\documentclass[preprint]{elsarticle}
\usepackage{graphicx}
\usepackage{natbib}
\usepackage{dcolumn}
\usepackage{bm}
\usepackage[utf8]{inputenc}
\usepackage{color}
\usepackage{epstopdf}
\DeclareGraphicsExtensions{.pdf,.eps}
\newcommand{\ket}[1]{\left| #1 \right>}
\newcommand{\bra}[1]{\left< #1 \right|}

\newcommand{\braket}[2]{\left< #1 | #2 \right>}

\begin{document}
\title{Bell states and entanglement dynamics on two coupled quantum molecules}
\author[rvt]{P. A. Oliveira}
\author[rvt]{L. Sanz}\ead{lsanz@infis.ufu.br}
\address[rvt]{Instituto de Física, Universidade Federal de Uberlândia, 38400-902,
Uberlândia-MG, Brazil.}
\begin{abstract}
This work provides a complete description of entanglement properties between electrons inside coupled quantum molecules, nanoestructures which consist of two quantum dots. Each electron can tunnel between the two quantum dots inside the molecule, being also coupled by Coulomb interaction.  First, it is shown that Bell states act as a natural basis for the description of this physical system, defining the characteristics of the energy spectrum and the eigenstates. Then, the entanglement properties of the eigenstates are discussed, shedding light on the roles of each physical parameters on experimental setup. Finally, a detailed analysis of the dynamics shows the path to generate states with a high degree of entanglement, as well as physical conditions associated with coherent oscillations between separable and Bell states.
\end{abstract}
\maketitle
\section{Introduction}
\label{sec:intro}
In the last two decades, the interest on the physical behavior of semiconductor materials has increased due to their potential applications in quantum computing~\cite{Kane98,Loss98}. In particular, semiconductor quantum dots have been proved to be an ideal candidate for the codification of quantum information. In these systems, a qubit can be defined using the charge~\cite{Hollenberg04} and the spin of the particle confined~\cite{Imamoglu99,Biolatti00,Biolatti02}, as well as excitonic states~\cite{Rolon10,Borges10}. An interesting system is a quantum molecule (QM) which is a nanoestructure consisting on two quantum dots separated by a barrier that can be tunneled by charged particles, as electrons~\cite{Kobayashi95,Tarucha96,Fujisawa98,Oosterkamp98}.

In 2003, Hayashi {\it{et al.}} demonstrated the coherent manipulation of a  charge qubit in a QM~\cite{Hayashi03}. The quantum nanoestruture is built by using metal gates for confining charges on a bidimensional electron gas in GaAs/AlGaAs. The same group implemented a quantum device with two QMs coupled electrostatically but isolated by conduction~\cite{Shinkai09}. In this system, the resonant tunneling current through each molecule is influenced by the charge state of the second molecule~\cite{Shinkai07}. More recently, researchers on the Quantum Device Lab (ETH Zurich) and Université of Sherbrooke (Québec, Canada) demonstrated successfully the coupling between the same type of device with a microwave resonator~\cite{Frey12}. The experimental data (together with a theoretical treatment) shows that the coupling between a quantum molecule and the resonator follows a Jaynes-Cummings type interaction. In another interesting work, quantum oscillations between three quantum states in a Si/SiGe double quantum dots are controlled by high-speed voltage pulses~\cite{Eriksson14}. Measurements of the transconductance permitted the observation of this oscillations between different coupled quantum states induced by two different pulse profiles. Both works show the potential applications of this kind of heterostructures on quantum information processing, opening the possibility of the definition and control of quantum bits.

From the theoretical point of view, Fujisawa \textit{et al.}~\cite{Fujisawa11} demonstrated that, in the experimental setup mentioned above, it is possible to implement some of the two-qubit gates, including the Bell gate at a very specific choice of physical parameters. Other theoretical works have explored aspects related to quantum correlations and decoherence~\cite{Fanchini10,Emary09}, also considering restricted physical conditions. Due the importance of the role of entanglement in quantum computation protocols, a complete analysis of the entanglement properties of this particular system, with potential application of quantum information processing, is crucial.

Here, it is developed a careful exploration of the entanglement properties of electrons in coupled QMs, considering the realistic experimental setup of Ref.~\cite{Shinkai09}, treated as a closed system. The results show that Bell states are the key behind the physical behavior of the electrons inside the coupled QMs. Also, it is demonstrated how the careful control of physical parameters as tunneling, electronic energy offsets (detunings), and Coulomb coupling, can be used to create highly entangled states. First, using analytical calculations together with numerical simulations, the characteristics of the energy spectrum and the eigenstates of the Hamiltonian are explored. The results show that the eigenstates correspond to Bell states under specific physical conditions. Second, the effect of physical parameters is mapped, focusing on the generation of highly entangled states. The ``beats" on the dynamics of entanglement are verified and explained as a competition between two different frequencies associated with the physical couplings. Physical requirements to obtain coherent oscillations between separable states and a specific Bell state are discussed. All results are obtained considering realistic values of physical parameters.

This paper is structured as follows. Section~\ref{sec:theory} contains the description of the theoretical model together with the definition of the measurement of entanglement, concurrence, used in this work. Section~\ref{sec:eigen} is devoted to explore the characteristics of the energy spectrum and the eigenstates from the point of view of entanglement properties. Section~\ref{sec:entdyn} is reserved to the discussion of dynamics, focussing in the obtention of Bell states using temporal evolution, considering the effects of tunneling and detuning between electronic states inside each QM. The work is summarized in Section~\ref{sec:summary}.

\section{Model}
\label{sec:theory}

The analysis performed here is based on an actual experimental setup with two coupled QMs~\cite{Shinkai09,Hayashi03}, each with two quantum dots separated by a potential barrier which can be tunneled by electrons~\cite{Kobayashi95,Tarucha96,Fujisawa98,Oosterkamp98}. The quantum dots for each QM are arranged horizontally, being coupled to a source and a drain of electrons, used to charge and discharge the QM.  Coulomb blockade and the physical design of the nanoestrutures guarantee that there is only one extra electron and, effectively, one electronic level for each quantum dot. Each QM can be treated theoretically as a two-level system: the electron can occupy the left (right) dot $\ket{L}$ ($\ket{R}$). Even if the two QMs are designed in order to inhibit intra-molecules tunneling, both subsystems are still coupled because of the Coulomb interaction between the electrons on each molecule. From the point of view of quantum information, this correspond to the implementation of two qubits, one on each QM, which are coupled in such a way that the physical system is bipartite.

This system is well modeled by the Hamiltonian~\cite{Shinkai09} written as
\begin{equation}
\hat{H}=\frac{1}{2}\sum_i\left(\varepsilon_{i}\sigma_{z}^{\left(i\right)}+\Delta_{i}\sigma_{x}^{\left(i\right)}\right)
+\frac{J}{4}\sigma_{z}^
{\left(1\right)}\otimes\sigma_{z}^{\left(2\right)}
\label{eq:H}
\end{equation}
where $\varepsilon_i$ parameters describe the energy offsets of the electronic levels, also known as detunings, and $\Delta_i$  are
tunneling rates inside each QM. Last term is the Coulomb interaction between electrons from different QM, which is quantified by $J$. Considering $\sigma_z^i=\ket{L}_{ii}\bra{L}-\ket{R}_{ii}\bra{R}$ and $\sigma_x^i=\ket{L}_{ii}\bra{R}+\ket{R}_{ii}\bra{L}$, the matrix form of Hamiltonian (\ref{eq:H}), written in the positional basis of the coupled molecules $\left\{\ket{M_{1}M_{2}}=\ket{LL},\ket{LR},\ket{RL},\ket{RR}\right\}$, is given by:
\begin{equation}
\hat{H}=\left(\begin{array}{cccc}
\frac{\varepsilon_{s}}{2}+\frac{J}{4} & \frac{\Delta_{2}}{2} & \frac{\Delta_{1}}{2} & 0\\
\frac{\Delta_{2}}{2} & \frac{\varepsilon_{d}}{2}-\frac{J}{4} & 0 & \frac{\Delta_{1}}{2}\\
\frac{\Delta_{1}}{2} & 0 & -\frac{\varepsilon_{d}}{2}-\frac{J}{4} & \frac{\Delta_{2}}{2}\\
0 & \frac{\Delta_{1}}{2} & \frac{\Delta_{2}}{2} & -\frac{\varepsilon_{s}}{2}+\frac{J}{4}\end{array}\right),
\label{eq:Hm}
\end{equation}
where $\varepsilon_{s}=\varepsilon_{1}+\varepsilon_{2}$ and $\varepsilon_{d}=\varepsilon_{1}-\varepsilon_{2}$. As measurement of entanglement, we use the concurrence, an efficient measurement for bipartite $2\otimes 2$ systems~\cite{Hill97,Wootters98}, which is defined as
\begin{equation}
C(\rho)=\mathrm{max}\left\lbrace 0,\lambda_1-\lambda_2-\lambda_3-\lambda_4\right\rbrace.
\end{equation}
Here $\lambda_{i}$ are the square roots of the eigenvalues, in decreasing order, of the matrix $R=\rho \tilde{\rho}$. The $\tilde{\rho}$ operator is obtained through the spin-flip operation of the complex conjugate of the density matrix of the $2\otimes 2$ system so $\tilde{\rho}=(\sigma_{y}^{A}\otimes\sigma_{y}^{B})\rho^*(\sigma_{y}^{A}\otimes\sigma_{y}^{B})$.
If $C(\rho)=0$, the state of the bipartite system is separable. On the other hand, when $C (\rho)=1$, the state is a maximally entangled state. In the following sections, the theoretical model and the entanglement measurement are used to explore the effects of physical parameters on both, the entanglement properties of eigenstates, and the entanglement dynamics.

\section{Entanglement properties of eigenvectors}
\label{sec:eigen}
Consider the Bell states given by~\cite{Nielsen}
\begin{eqnarray}
\ket{\Psi_{\pm}}&=&\frac{1}{\sqrt{2}}\left(\ket{RL}\pm \ket{LR}\right),\nonumber\\
\ket{\Phi_{\pm}}&=&\frac{1}{\sqrt{2}}\left(\ket{RR}\pm \ket{LL}\right).
\label{eq:Bell}
\end{eqnarray}
These are eigenvectors of the Hamiltonian (\ref{eq:Hm}) when $\Delta_1=\Delta_2=0$ and $\epsilon_1=\epsilon_2=0$ and also form a complete basis for $4\otimes 4$ Hilbert space. Using the ordering $O_B=\left\{\ket{\Psi_-},\ket{\Phi_-},\ket{\Psi_+},\ket{\Phi_+}\right\}$, the Hamiltonian (\ref{eq:Hm}) becomes:
\begin{equation}
\hat{H}_{B}=\left(\begin{array}{cccc}
-\frac{J}{4} & \Delta_{-} & -\frac{\varepsilon_{d}}{2} & 0\\
\Delta_{-}& \frac{J}{4} & 0 &-\frac{\varepsilon_{s}}{2}\\
-\frac{\varepsilon_{d}}{2} & 0 & -\frac{J}{4} & \Delta_{+}\\
0 & -\frac{\varepsilon_{s}}{2}& \Delta_{+}  & \frac{J}{4}\end{array}\right)
=\left(\begin{array}{cc}
{M_-} & {D}\\
{D}&{M_+}\end{array}\right),
\label{eq:Hbell}
\end{equation}
where $\Delta_{\pm}=(\Delta_1\pm\Delta_2)/2$. Written in the Bell basis, it becomes evident that tunneling and Coulomb couplings split the Hilbert space in two $2\times 2$ subspaces, described by matrices ${M_{\pm}}$, with the matrix ${D}$ coupling both of them. Rewriting Hamiltonian (\ref{eq:H}) using the Bell basis is advantageous to explore the rich entanglement behavior of the system: it separates the effect of tunneling $\Delta_-(\Delta_+)$, which is responsible for the oscillations within 2-fold subspaces $\ket{\Psi_-}(\ket{\Psi_+})$ and $\ket{\Phi_-}(\ket{\Phi_+})$, from the effect of detunings $\varepsilon_{s}$ and $\varepsilon_{d}$, responsible for inter-subspaces mixing. This new approach is ideal to study the entanglement properties of eigenstates using both, analytical and numerical calculations.
\subsection{Analytical solution at $\epsilon_1=\epsilon_2=0$}
An analytical solution is obtained by considering the resonant electronic levels so $\epsilon_1=\epsilon_2=0$. Notice that, for this condition, the $2\times 2$ matrix ${D}$ in Eq.(\ref{eq:Hbell}) is null so the four Hamiltonian eigenvalues are given by
\begin{equation}
E_{\mp}^{\Delta_{\mp}}=\mp\frac{\sqrt{J^2+16\Delta^2_{\mp}}}{4},
\label{eq:energiesres}
\end{equation}
with the corresponding eigenstates written as
\begin{eqnarray}
\ket{\psi_-^{\Delta_{\mp}}}&=&\Gamma_-^{\Delta_{\mp}}\left(\ket{\Psi\mp}+\Omega_-^{\Delta_{\mp}}\ket{\Phi\mp}\right)\nonumber\\
\ket{\psi_+^{\Delta_{\mp}}}&=&\Gamma_+^{\Delta_{\mp}}\left(\ket{\Phi\mp}+\Omega_+^{\Delta_{\mp}}\ket{\Psi\mp}\right).
\label{eq:statesres}
\end{eqnarray}
Here, the new quantity $\Omega_{\mp}^{\Delta_{\mp}}$ is defined as
\begin{equation}
\Omega_{\mp}^{\Delta_{\mp}}=\mp\frac{J-\sqrt{J^2+16\Delta^2_{\mp}}}{4\Delta_{\mp}},
\end{equation}
while $\Gamma=1/\sqrt{1+\Omega^2}$ is the normalization term. The eigenstates are now a superposition of two elements on Bell basis which depends on the tunneling coupling. It is straightforward to check the validity of this solution by considering the limiting case of tunneling suppression ($\Delta_1=\Delta_2=0$). For this condition, the function $\Omega_{j}^{\Delta_{i}}$ ($j=\mp$) goes asymptotically to zero when $\Delta_{i}\rightarrow0$ $(i=\mp)$ and, consequently, $\Gamma\rightarrow1$ so the set given by $\left\{\ket{\psi_-^{0}},\ket{\psi_+^{0}},\ket{\psi_-^{0}},\ket{\psi_+^{0}}\right\}$ coincides with $\left\{\ket{\Psi_-},\ket{\Phi_-},\ket{\Psi_+},\ket{\Phi_+}\right\}$.

The analytical solutions, Eqs.(\ref{eq:energiesres},\ref{eq:statesres}), are useful to explore the physical consequences of fixing equal tunneling rates ($\Delta_2=\Delta_1=\Delta$). Under this condition, the matrix $M_-$ in Hamiltonian (\ref{eq:Hbell}) becomes diagonal, while $M_+$ remains non-diagonal with $\Delta_+=\Delta$. Depending on the relations between parameters $\varepsilon_1$ and $\varepsilon_2$, the system falls into one of the following behaviors:
\begin{itemize}
\item At full resonance, when $\varepsilon_2=\varepsilon_1=0$, the eigenstates correspond to Eq.(\ref{eq:statesres}), with two of them corresponding to the Bell states $\ket{\Psi_-}$ and $\ket{\Phi_-}$ with energies $-J/4$ and $J/4$ respectively. The others are given by $\ket{\psi_{\mp}^{\Delta}}$ with energies $E_{\mp}^{\Delta}$.
\item Considering $\varepsilon_2=\varepsilon_1$, which means $\varepsilon_d=0$, the eigenstate $\ket{\psi_-^{0}}$ remains as $\ket{\Psi_-}$, a maximum entangled state, while $\ket{\Phi_-}$ couples with subspace ``+" by the action of detuning $\varepsilon_s$.
\item If  $\varepsilon_2=-\varepsilon_1$, then $\varepsilon_s=0$, the eigenstate $\ket{\psi_+^{0}}$ becomes the Bell state $\ket{\Phi_-}$ while the other three eigenstates are coupled.
\end{itemize}

\subsection{Numerical results}
\label{subsec:numerical}
In this subsection, the entanglement properties of the coupled QMs are explored in a wider scenario: first, the action of detunings is studied when the system has equal tunneling rates, $\Delta_-=0$. Then, the effects of both tunneling terms, $\Delta_+$ and $\Delta_-$, are quantified.
\subsubsection{Equal tunneling rates}
\begin{figure}[tbp]
\includegraphics[scale=.7]{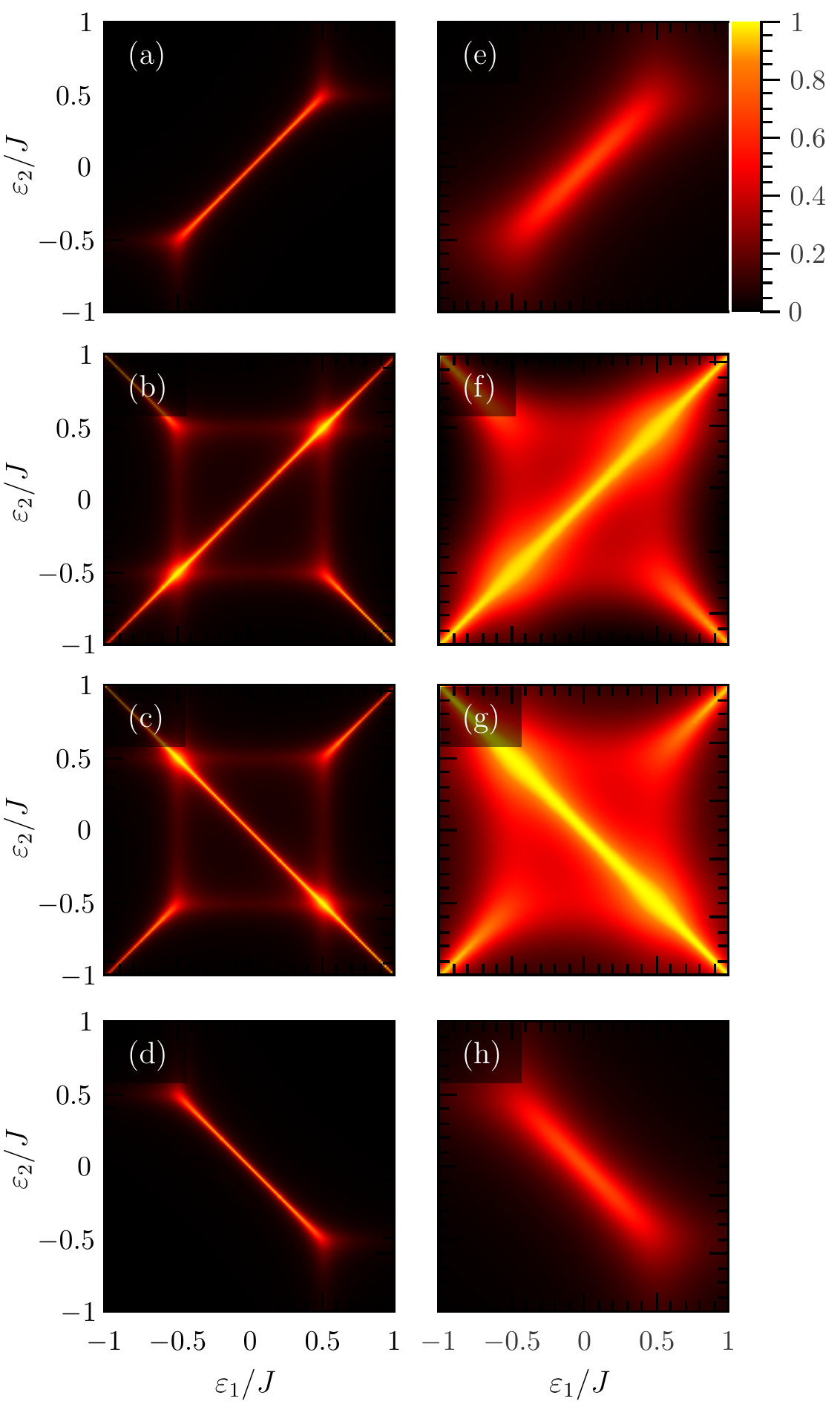}
\caption{Concurrence of the eigenstates of Hamiltonian (\ref{eq:H}) as a function of $\varepsilon_1$ and $\varepsilon_2$ with $J=25\mu$eV. For $\Delta_1=\Delta_2=\Delta=J/16$: (a) $\ket{0}$, (b)$\ket{1}$, (c) $\ket{2}$ and (d) $\ket{3}$. For $\Delta_1=\Delta_2=\Delta=J/4$: (e) $\ket{0}$, (f)$\ket{1}$, (g) $\ket{2}$ and (h) $\ket{3}$.}
\label{fig:concdelta1}
\end{figure}

Fig.~\ref{fig:concdelta1} shows the numerical results for concurrence of the eigenstates of Hamiltonian (\ref{eq:Hm}) as a function of detunings $\varepsilon_i$, considering equal tunneling rates ($\Delta_-=0$ and $\Delta_+=\Delta$). In this calculations, two different choices of tunneling parameters are used: $\Delta=J/16$ (weak tunneling), Figs.~\ref{fig:concdelta1}(a-d), and $\Delta=J/4$ (strong tunneling), Figs.~\ref{fig:concdelta1}(e-h).
The Coulomb coupling is fixed at $J=25$ $\mu$eV, as reported in experiments~\cite{Shinkai09,Hayashi03}. For all cases, the eigenstates are shown in ascending order of energy so $\left\{\ket{0},\ket{1},\ket{2},\ket{3}\right\}$ are ground state and the first, second and third excited states respectively.

First, the analytical results for the three types of conditions discussed above are confirmed with the numerical calculations: at full resonance, eigenstate $\ket{0}$ and $\ket{3}$ correspond to $\ket{\psi_{-}^{\Delta}}$ and $\ket{\psi_{+}^{\Delta}}$ respectively while $\ket{1}=\ket{\Psi_-}$ and $\ket{2}=\ket{\Phi_-}$ are maximally entangled states, fact confirmed by checking the coefficients of the eigenstates calculated by the numerical simulation and shown by the bright yellow line over the condition $\varepsilon_2=\varepsilon_1$ in Figs.~\ref{fig:concdelta1}(b) and (f). For the condition $\varepsilon_d=0$, $\ket{1}$ remains in $\ket{\Psi_-}$, while at $\varepsilon_2=-\varepsilon_1$, the second excited state $\ket{2}=\ket{\Phi_-}$, conditions associated with yellow lines seen in Figs.~\ref{fig:concdelta1}(c) and (g).

Out of the three resonant cases, the value of concurrence has an abrupt decrease due to the enhancement of coupling between subspaces associated with $M_{\pm}$. The eigenstates $\ket{0}$ and $\ket{3}$ show a lower entanglement degree, if compared with $\ket{1}$ and $\ket{2}$, with some differences that can be seen in Fig.~\ref{fig:concdelta1}(a) and Fig.~\ref{fig:concdelta1}(e), for state $\ket{0}$, and Fig.~\ref{fig:concdelta1}(d) and Fig.~\ref{fig:concdelta1}(h), for state $\ket{3}$: at the full resonant condition and weak tunneling rate, $\ket{0}$ remains as $\ket{\psi_{-}^{\Delta}}$ while $\ket{3}=\ket{\psi_{+}^{\Delta}}$, both being superpositions with high concurrence value given by $C(\rho)\simeq 0.9$. Then, following the condition $\varepsilon_2=\varepsilon_1$ in Fig.~\ref{fig:concdelta1}(a) and $\varepsilon_2=-\varepsilon_1$ in Fig.~\ref{fig:concdelta1}(d), the value of concurrence decreases being $C(\rho)\simeq 0.5$ at the two points corresponding to $|\varepsilon_i|=J/2$. The decreasing value of concurrence is the signature of coupling between subspace $M_{+}$ with one of the Bell states: for $\ket{0}$, the coupled states are $\ket{\psi_{-}^{\Delta}}$ and $\ket{\Phi_-}$ while in state $\ket{3}$ the coupling is between $\ket{\psi_{+}^{\Delta}}$ and $\ket{\Psi_-}$. It is expected that the three choices for $\varepsilon_i$ which favored highly entangled states as eigenstates of the system affect the behavior of the entanglement dynamics of this system.

For values of $\varepsilon_i$ out of resonance conditions, the four Bell states are coupled by both, the tunneling and detunings, and the mixing of basis elements explains the lower value of concurrence. Nevertheless, by comparing the two cases of tunneling, the values of concurrence increase when $\Delta$ increases, fact explained because tunneling favored the coupling between internal states on each subspaces, which forms a superposition less sensitive to the action of detuning. Still, the numerical calculations show that the high degree of entanglement of superpositions $\ket{\psi_+^{\Delta_{\mp}}}$, with $C\sim0,97$ obtained for $\Delta=J/16$ at full resonance condition, Figs.~\ref{fig:concdelta1}(a,d), decreases as tunneling rate increases, being $C\sim0,70$ for the same resonant condition in Figs.~\ref{fig:concdelta1}(e,h).

It is interesting to note that, concerning detunings, there are two different set of parameters: the first set is defined by condition $\varepsilon_{i}\in\left[-J/2,J/2\right]$ where sub-spaces $M_+$ and $M_-$ are coupled with each other. For the second set of values, it is verify that the detunings have the effect of exchanging the value of the energies associated with $\ket{1}$ and $\ket{2}$ for even and odd resonances, still coupling the four Bell states out of any resonant conditions. For higher values of $\Delta$, Figs.~\ref{fig:concdelta1}(e-h), the increase of value of tunneling rate has the effect of enhancing the coupling inside the subspace associated with $M_+$, which causes the increase of the ``area" with high values of concurrence, $0.45<C<1$ as can be easily check by comparing (a) and (e) in the figure. Thus, the set of conditions for entangled states for $\Delta=J/16$ expands covering a wide region, although the expansion happens around the fixed critical points $\varepsilon_i=\pm J/2$.
\subsubsection{Different tunneling rates}
\begin{figure}[tbp]
\includegraphics[scale=0.7]{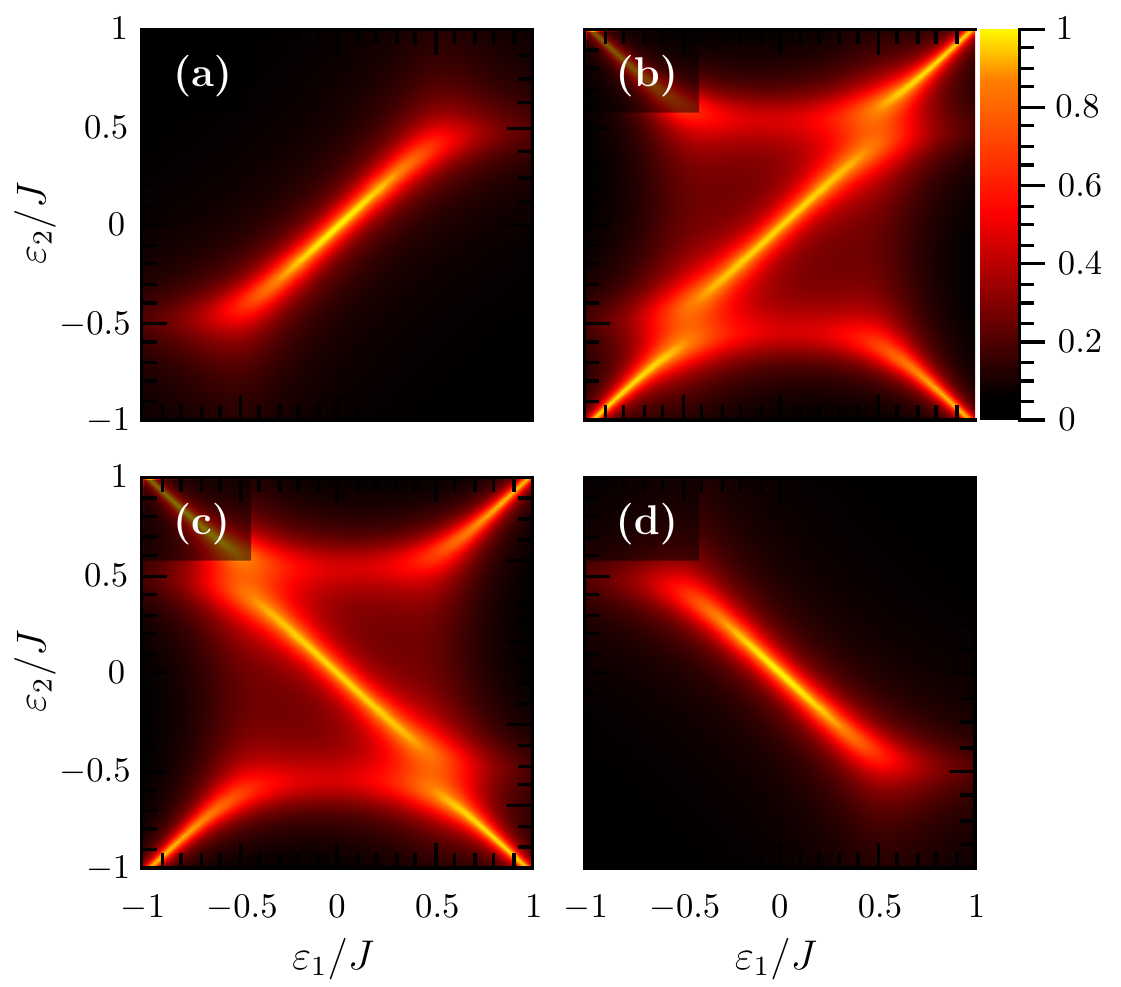}
\caption{Concurrence of the eigenstates of Hamiltonian (\ref{eq:H}) as a function of $\varepsilon_1$ and $\varepsilon_2$ considering $\Delta_1=J/4$ and $\Delta_2=J/8$ for $J=25\mu$eV: (a) $\ket{0}$, (b)$\ket{1}$, (c) $\ket{2}$ and (d) $\ket{3}$. }
\label{fig:concdelta2}
\end{figure}

Fig.~\ref{fig:concdelta2} shows the behavior of concurrence for $\Delta_1\neq\Delta_2$, considering $\Delta_1=J/4$ and $\Delta_2=J/8$. In this case, the eigenstates of the system are not Bell states at any condition of resonance. Inside the region with $\varepsilon_{i}\in\left[-J/2,J/2\right]$, it is observed the effect of the coupling between subspaces $M_+$ and $M_-$: around the conditions $\varepsilon_2=\varepsilon_1$ and $\varepsilon_2=-\varepsilon_1$, there is a region with higher entanglement degree for all eigenstates, although the concurrence is higher for the states associated with $M_-$, $\ket{1}$ and $\ket{2}$. This is explained because the effective tunneling $\Delta_-$ is weaker than $\Delta_+$. Out of these limits, there is a region with a high degree of entanglement for eigenstates $\ket{1}$ and $\ket{2}$, a new feature if compared with Fig.~\ref{fig:concdelta1}. That means different tunneling rates expand the range of physical conditions for obtaining entangled states with a high degree of entanglement ($C>0.7$), although Bell states are not longer attained.

\section{Dynamical generation of entangled states}
\label{sec:entdyn}

A path for creation of highly entangled states is the use of dynamical evolution. First requirement is to prepare the system in an initial state. At experiments, it is possible to manipulate the energy offsets to set a specific configuration on couple QMs. The process is highly controlled~\cite{Shinkai09} and $\ket{\Psi(0)}$ can be fixed as one of the states $\left\{\ket{LL},\ket{LR},\ket{RL},\ket{RR}\right\}$ of the positional basis. Tunneling parameters are defined by construction of the nanoestructure, also by additional gates that inhibit or enhance coherent tunneling. The setup is open to manipulation of the detuning parameters $\varepsilon_i$. The readout of the state after the evolution is performed manipulating the bias on QMs and measuring the contribution of one electron to the current~\cite{Shinkai09}. Inside this context, the effects of the physical couplings on quantum dynamics can be studied using analytical and numerical simulations.

\subsection{Tunneling effect on entanglement dynamics}
\label{subsec:tundyn}
\begin{figure}[tbp]
\includegraphics[scale=0.5]{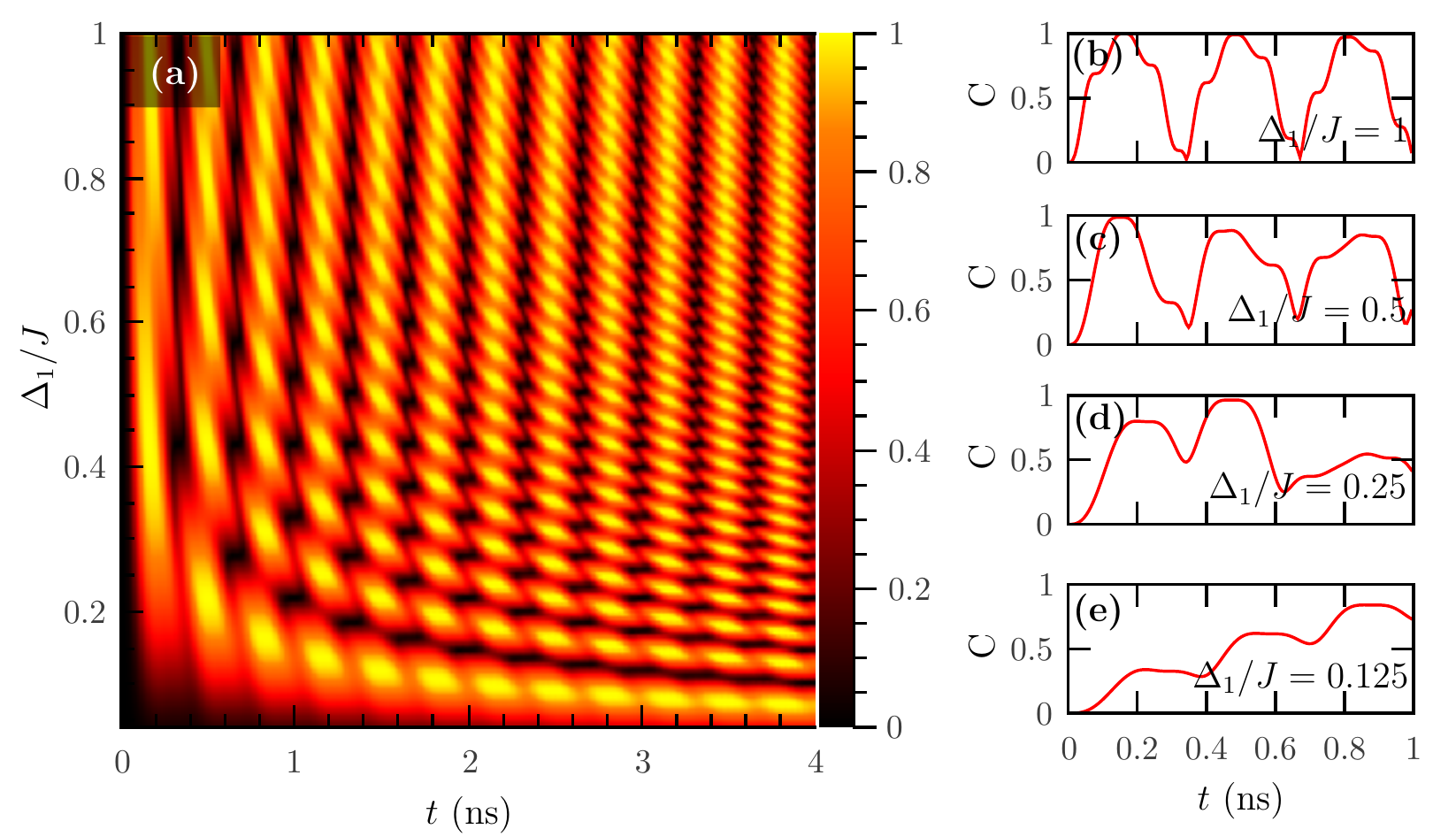}
\caption{Concurrence dynamics considering $\ket{\Psi(0)}=\ket{RL}$ with $J=25$ $\mu$eV. (a) Concurrence as a function of time and $\Delta_1/J$ considering $\Delta_1=\Delta_2$ for $0<t<3$ ns; (b-e) Concurrence evolution for $0<t<1$ ns considering specific choices for $\Delta/J$ ratio.}
\label{fig:deltadyn}
\end{figure}

The first goal is to analyze the effect of tunneling rates, once the results obtained in Section~\ref{sec:eigen} show the relevance of those coupling in the characteristics of the energy spectrum. A numerical approach provides insights about the dynamical preparation of highly entangled states. In Fig.~\ref{fig:deltadyn}(a), the obtained numerical results of concurrence are shown as a function of time and $\Delta_1/J$ ratio considering $\ket{\Psi(0)}=\ket{RL}$, $J=25$ $\mu$eV, $\Delta_1=\Delta_2$ and resonance condition given by $\varepsilon_2=\varepsilon_1=0$. It is verified that the tunneling coupling has a strong effect on entanglement dynamics: the evolved state goes from separable (dark) to a highly entangled state (bright) after the evolution time $t_e$ which depends on $\Delta_1/J$. This can be verified by comparing the four profiles in Figs.~\ref{fig:deltadyn}(b-e), noticing that the time when the concurrence has a local maximum at times which depend on the value of $\Delta_1/J$, also shown in Figs.~\ref{fig:deltadyn}(b-e). It is interesting to observe that the oscillations are not fully periodical. This kind of behavior is a signature of the dynamical competition between different couplings, each with a characteristic frequency. This is better seen in Figs.~\ref{fig:deltadyn}(b-e), where the patterns resemble the ``beats" obtained with the superposition of two harmonic oscillators.

\begin{figure}[tbp]
\includegraphics[scale=0.5]{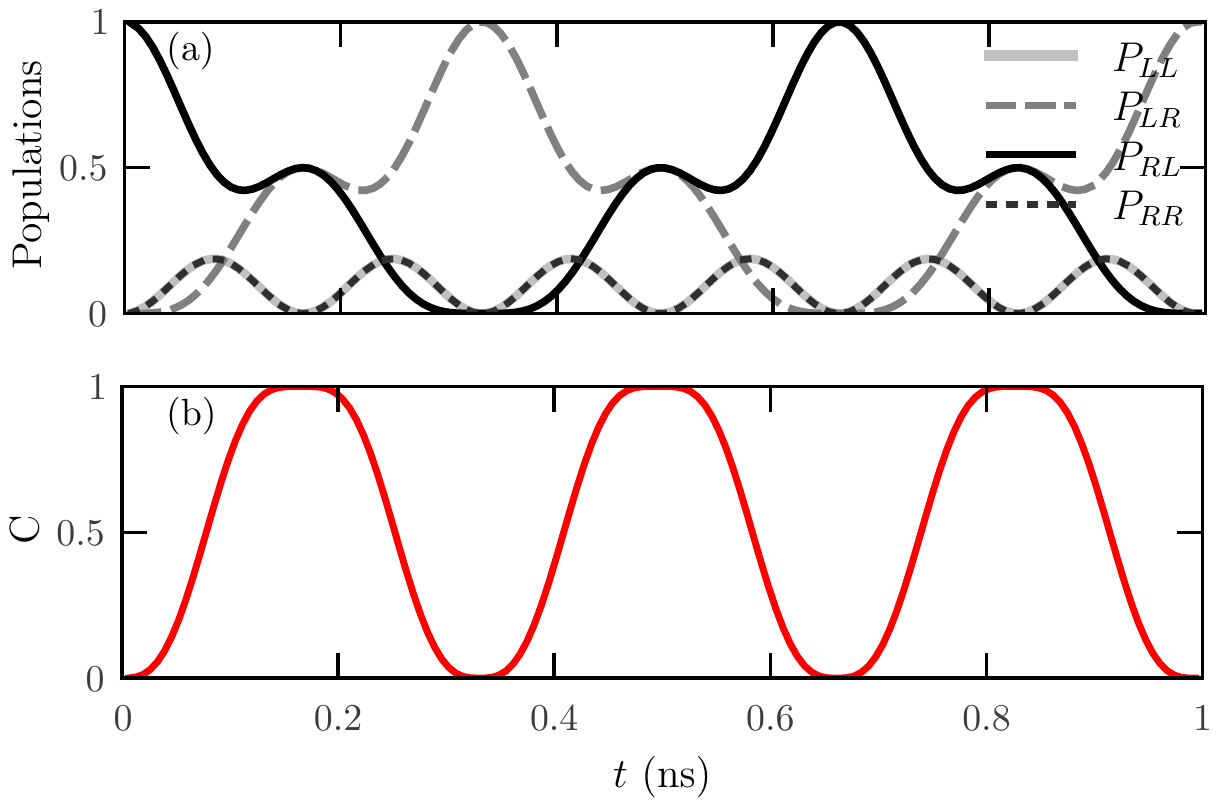}
\caption{Populations and Concurrence dynamics for initial state given by $\ket{\Psi(0)}=\ket{RL}$. Here  $J=25$ $\mu$eV, $\varepsilon_2=\varepsilon_1$ and $\Delta_1=\Delta_2$ being $\Delta_1/J=\sqrt{3}/4$ which fulfills condition given by Eq.(\ref{eq:cohbellcon}) considering $n=m=1$. (a) Populations of the four positional states $\ket{M_1M_2}$ as a function of time; (b) Periodical oscillation of Concurrence.}
\label{fig:bellosc}
\end{figure}

To shed more light on the physical origins behind this concurrence beats, the time-dependent Schrödinger equation is solved considering the full resonant condition, $\varepsilon_1=\varepsilon_2=0$. This can be done using the previous results for eigenergies, Eq.(\ref{eq:energiesres}), and eigenstates, Eq.(\ref{eq:statesres}). If $\ket{\Psi(0)}=\ket{RL}$, it is straightforward to obtain the evolved state $\ket{\Psi(t)}$ and then the populations for the states of computational basis, $P_{ij}=|\braket{ij}{\Psi(t)}|^2$. The last are written as:
\begin{eqnarray}
P_{RL}&=&\frac{1}{4}\Bigg[\left(\frac{J}{\beta_+}\sin{\Omega_+t}+\frac{J}{\beta_-}\sin{\Omega_-t}\right)^2\nonumber\\
&&+\left(\cos{\Omega_+t}+\cos{\Omega_-t}\right)^2\Bigg],\nonumber\\
P_{RL}&=&\frac{1}{4}\Bigg[\left(\frac{J}{\beta_+}\sin{\Omega_+t}-\frac{J}{\beta_-}\sin{\Omega_-t}\right)^2\nonumber\\
&&+\left(\cos{\Omega_+t}-\cos{\Omega_-t}\right)^2\Bigg],\nonumber\\
P_{RR}&=&4\left[\frac{\Delta_+}{\beta_+}\sin{\Omega_+t}+\frac{\Delta_-}{\beta_-}\sin{\Omega_-t}\right]^2,\nonumber\\
P_{LL}&=&4\left[\frac{\Delta_+}{\beta_+}\sin{\Omega_+t}-\frac{\Delta_-}{\beta_-}\sin{\Omega_-t}\right]^2,\\
\label{eq:pop}
\end{eqnarray}
with the new physical parameters $\Omega{\pm}$ defined as
\begin{equation}
\Omega_{\pm}=\frac{\beta_{\pm}}{4}=\frac{\sqrt{J^2+16\Delta_{\pm}^{2}}}{4}.
\end{equation}
These two frequencies are behind the concurrence beats in Fig.~\ref{fig:bellosc}. Because $\Omega_{\pm}$ are functions of the tunneling parameters $\Delta_{\pm}$, the mechanism behind the rich entanglement dynamics shown in Fig.~\ref{fig:deltadyn} is the coupling between Bell states inside the subspaces $M_{\pm}$. That means the action of the tunneling can be used to manipulate the obtention of highly entangled states.

There is an exact condition for coherent periodic oscillations between separable and Bell states: notice that states $\ket{LL}$ and $\ket{RR}$ are not populated for evolution times corresponding to
\begin{equation}
t_{e}=\frac{n\pi}{\Omega_+},
\label{eq:t1}
\end{equation}
with $n$ being any integer number. At these specific times, the expressions $P_{LR}$ and $P_{RL}$ on Eq.(\ref{eq:pop}) reduce to
\begin{eqnarray}
P_{RL}&=&\frac{1}{4}\left\{\sin^2{\Omega_-t_e}+\left[(-1)^n+\cos{\Omega_-t_e}\right]^2\right\},\nonumber\\
P_{RL}&=&\frac{1}{4}\left\{\sin^2{\Omega_-t_e}+\left[(-1)^n-\cos{\Omega_-t_e}\right]^2\right\}.\nonumber\\
\label{eq:popLRRL}
\end{eqnarray}
In this way, the behavior of the function $\cos(\Omega_+ t_e)$ defines the condition for generation of Bell states $\ket{\Psi_{\pm}}$ in Eq.(\ref{eq:Bell}).
If its value is zero, the population of both states goes to $1/2$, indicating that the system goes to a Bell state. The evolution time is also given by
\begin{equation}
t_e=\frac{m\pi}{2\Omega_-},\quad m\;\mathrm{odd}.
\label{eq:tevolv}
\end{equation}
Comparing Eq.(\ref{eq:t1}) and Eq.(\ref{eq:tevolv}), both evolution times are equal if condition $\frac{n}{m}=\frac{\Omega_+}{2\Omega_-}$ is satisfied. In term of coupling parameters, the condition is equivalent to
\begin{equation}
\frac{\Delta_1}{J}=\frac{1}{4}\sqrt{4\frac{n^2}{m^2}-1},
\label{eq:cohbellcon}
\end{equation}
which has a real solution for $m<2n$. The populations and concurrence dynamics are shown in Fig.~\ref{fig:bellosc}, for physical parameters which follow the conditions given by Eq.(\ref{eq:cohbellcon}). Notice that, at , the initial state $\ket{RL}$ exchanges population with $\ket{LR}$, to return again after two periods of oscillation. At times defined by Eq.(\ref{eq:tevolv}), at times when the evolved state is a separable one, Bell states with concurrence value $C=1$ are obtained as a superposition of the $\ket{RL}$ and $\ket{LR}$ states. Numerical and analytical results obtained by considering the other three states of positional basis $\ket{M_1 M_2}$, not shown here, establish that the initial states given by $\ket{LR}$ or $\ket{RL}$ evolve to Bell states $\ket{\Psi_{\pm}}$, while Bell states $\ket{\Phi_{\pm}}$ are dynamically created by setting $\ket{LL}$ and $\ket{RR}$ as initial states.

\subsection{Effects of detuning on concurrence dynamics}
\label{subsec:epsdyn}
The analysis of the concurrence behavior of eigenstates is a guide to check the effects of detunings given that, additionally to the full resonant condition, highly entangled states becomes eigenstates of the system at conditions given by $\varepsilon_2=\varepsilon_1$ and $\varepsilon_2=-\varepsilon_1$. When thinking about dynamics, the presence of such eigenstates at these specific conditions could favor the dynamical generation of Bell states.

\begin{figure}[hb]
\includegraphics[scale=0.55]{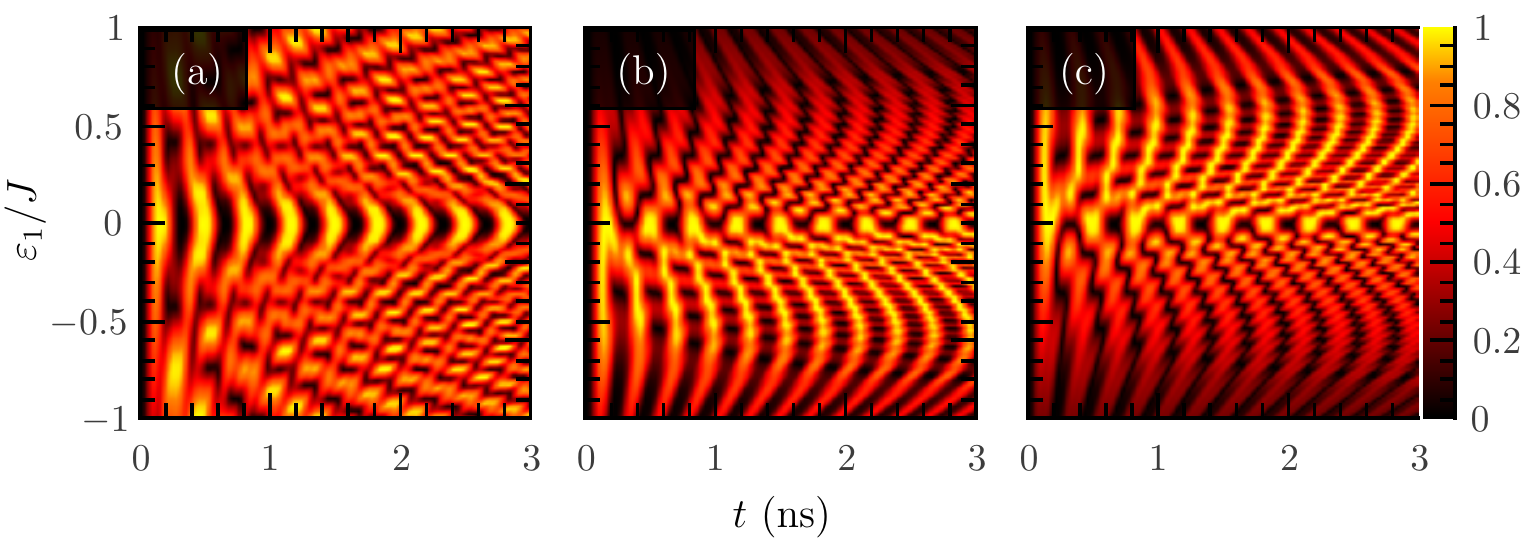}
\caption{Three patterns for concurrence dependence of detunings $\varepsilon_i$ and time, for tunneling couplings ($\Delta_1=\Delta_2$) and Coulomb coupling $J$ so $\Delta_1/J=\sqrt{3}/4$. (a) $\ket{\Psi(0)}=\left\{\ket{RL},\ket{LR}\right\}$ for $\varepsilon_2=\varepsilon_1$ and $\ket{\Psi(0)}=\left\{\ket{LL},\ket{RR}\right\}$ for $\varepsilon_2=-\varepsilon_1$; (b) $\ket{\Psi(0)}=\ket{RL}$ considering $\varepsilon_2=\varepsilon_1$ and $\ket{\Psi(0)}=\ket{LL}$ for condition $\varepsilon_2=-\varepsilon_1$; (c) $\ket{\Psi(0)}=\ket{LR}$ considering $\varepsilon_2=\varepsilon_1$ and $\ket{\Psi(0)}=\ket{RR}$ for $\varepsilon_2=-\varepsilon_1$.}
\label{fig:detudyn}
\end{figure}

To obtain a full insight of the effects of detuning, numerical simulations of twelve different situations, considering as initial states the four elements $\ket{M_1M_2}$ of positional basis and the conditions $\varepsilon_2=\varepsilon_1$ and $\varepsilon_2=-\varepsilon_1$ for $\varepsilon_1\in\left[-J,J\right]$, chosen based on the analysis of spectrum as discussed in Sec.~\ref{sec:eigen}. The results are shown in Fig.~\ref{fig:detudyn}. Each of the twelve physical situations fall in one of the three different patterns as follow:
\begin{itemize}
\item Fig.~\ref{fig:detudyn}(a):
\begin{enumerate}
\item $\ket{\Psi(0)}=\ket{RL}$ and $\ket{LR}$ with $\varepsilon_2=\varepsilon_1$;
\item  $\ket{\Psi(0)}=\ket{LL}$ and $\ket{RR}$ with $\varepsilon_2=-\varepsilon_1$
\end{enumerate}
\item Fig.~\ref{fig:detudyn}(b):
\begin{enumerate}
\item $\ket{\Psi(0)}=\ket{RL}$ and $\varepsilon_2=\varepsilon_1$;
\item $\ket{\Psi(0)}=\ket{LL}$ and $\varepsilon_2=-\varepsilon_1$.
\end{enumerate}
\item Fig.~\ref{fig:detudyn}(c):
\begin{enumerate}
\item $\ket{\Psi(0)}=\ket{LR}$ with $\varepsilon_2=-\varepsilon_1$;
\item $\ket{\Psi(0)}=\ket{RR}$ with $\varepsilon_2=\varepsilon_1$.
\end{enumerate}
\end{itemize}
Based on these results, we conclude that choices of detunings out of the full resonant condition do not break the link between the sets $\left\{\ket{RL},\ket{LR}\right\}$ and $\left\{\ket{LL},\ket{RR}\right\}$, each with a specific detuning condition associated with oscillations between a product of separable states and the corresponding Bell states. This is closely related to the characteristics of the eigenstates discussed in Sec.~\ref{sec:eigen}: the formation of Bell state $\ket{\Psi_{-}}$ ($\ket{\Phi_-}$), a superposition of $\ket{RL}$ and $\ket{LR}$ ($\ket{RR}$ and $\ket{LL}$), is favored by the condition $\varepsilon_2=\varepsilon_1$ ($\varepsilon_2=-\varepsilon_1$).

Nevertheless, the full resonant condition optimizes the coherent oscillations for all cases, as shown by the yellow regions between $-J/4<\varepsilon_1<J/4$ in Fig.~\ref{fig:detudyn}(a). Out of this choice of parameters, the system falls into one of the asymmetrical patterns shown in Fig.~\ref{fig:detudyn}(b) and Fig.~\ref{fig:detudyn}(c). Depending on the initial state, the creation of high entangled states is favored by negative or positive values of $\varepsilon_1$. One feature common in the three patterns is the change of the behavior of the oscillations at values of $|\varepsilon_1|>J/2$. This is also explained by the characteristic of eigenstates, which shows a different behavior for values out of $\varepsilon_i\in[-J/2,J/2]$, as show in Fig.~\ref{fig:concdelta1} and Fig.~\ref{fig:concdelta2}.

\section{Conclusions and perspectives}
\label{sec:summary}
Here, it is presented a careful analysis of entanglement properties of two electrons inside coupled QMs, focusing on the characteristics of the eigenstates and the quantum dynamics. The concurrence is used as entanglement measurement, once there are two coupled qubits defined in this particular system. First, it is demonstrated that the physical system is intrinsically linked to Bell (maximally entangled) states. Although the primary coupling connecting the QMs is the Coulomb coupling ($J$) between electrons, the entanglement properties of the eigenstates depends strongly of the tunneling coupling ($\Delta_i$) and the energy offsets ($\varepsilon_i$), also called as detunings . Specifically, the tunneling splits the $4\times 4$ Hilbert space in two $2\times 2$ subspaces of Bell states, while the role of detunings is to mix the two subspaces. The physical conditions for obtaining highly entangled eigenstates are fully described by analytical and numerical calculations, which includes specific parameter choices for obtaining Bell states as eigenstates.

Second,  it is explored the dynamics of the system in order to create highly entangled states. The results provide a full characterization of the role of tunneling and detuning. When ressonant electronic levels ($\varepsilon_i=0$) are considered, it is shown that the tunneling coupling is the key behind the controlled generation of Bell states. The apparition of ``beats" in the dynamics is caused by competition between two different frequencies, which are functions of both, Coulomb and tunneling couplings. This beats can become coherent oscillations between positional states and a Bell state for a specific choice of ratio between tunneling and Coulomb couplings. Out of the resonant condition, this coherent oscillations are very sensitive to changes on the values of detuning.

As future works, we want to search for a proposal of experimental measurement of the degree of entanglement in this system. In a theoretical proposal Emary~\cite{Emary09} demonstrated that the process of decay of the qubits on coupled quantum molecules, in the long-time limit, induces rotations on each qubit as well as the direct decay. Based on this fact, the author propose a correlator which can be used as measurement of entanglement through a CHSH inequality\cite{CHSHpaper}. The correlator is a function of concurrence as defined in this work. One of our perspectives for future work is to explore the definition of a different entanglement witness, looking for signatures of entangled states in the electronic current, generated by the system when pumped and drained. Another future research is the understanding the effects of dephasing and decoherence mechanisms over general dynamic, as well as the search for robust states.

\section{Acknowledgments}
\label{sec:ack}
The authors want to thank Marcelo de França Santos for useful discussion and the critical reading of manuscript. We also thank the referee for the helpful comments. This work was supported by CNPq, Capes, FAPEMIG and the Brazilian National Institute of Science and Technology of Quantum Information
(INCT-IQ).
\bibliographystyle{elsarticle-num}
%\bibliography{refoliveira13}

\begin{thebibliography}{10}
\expandafter\ifx\csname url\endcsname\relax
  \def\url#1{\texttt{#1}}\fi
\expandafter\ifx\csname urlprefix\endcsname\relax\def\urlprefix{URL }\fi
\expandafter\ifx\csname href\endcsname\relax
  \def\href#1#2{#2} \def\path#1{#1}\fi

\bibitem{Kane98}
B.~E. Kane, A silicon-based nuclear spin quantum computer, Nature 393 (1995)
  137--137.

\bibitem{Loss98}
D.~Loss, P.~DiVincenzo, Quantum computation with quantum dots, Phys. Rev. A 57
  (1998) 120--126.

\bibitem{Hollenberg04}
L.~C.~L. Hollenberg, A.~S. Dzurak, C.~Wellard, A.~R. Hamilton, D.~J. Reilly,
  G.~J. Milburn, R.~G. Clark, Charge-based quantum computing using single
  donors in semiconductors, Phys. Rev. B 69 (2004) 113301.

\bibitem{Imamoglu99}
A.~Imamo\u{g}lu, D.~D. Awschalom, G.~Burkard, D.~P. {DiVincenzo}, D.~Loss,
  M.~Sherwin, A.~Small, Quantum information processing using quantum dot spins
  and cavity {QED}, Phys. Rev. Lett. 83 (1999) 4204--4207.

\bibitem{Biolatti00}
E.~Biolatti, R.~C. Iotti, P.~Zanardi, F.~Rossi, Quantum information processing
  with semiconductor macroatoms, Phys. Rev. Lett. 85 (2000) 5647--5650.

\bibitem{Biolatti02}
E.~Biolatti, I.~D'Amico, P.~Zanardi, F.~Rossi, Electro-optical properties of
  semiconductor quantum dots: Application to quantum information processing,
  Phys. Rev. B 65 (2002) 075306.

\bibitem{Rolon10}
J.~E. Rolon, S.~E. Ulloa, Coherent control of indirect excitonic qubits in
  optically driven quantum dot molecules, Phys. Rev. B 82 (2010) 115307.

\bibitem{Borges10}
H.~S. Borges, L.~Sanz, J.~M. Villas-B\^oas, A.~M. Alcalde, Robust states in
  semiconductor quantum dot molecules, Phys. Rev. B 81 (2010) 075322.

\bibitem{Kobayashi95}
Q.~Xie, A.~Madhukar, P.~Chen, N.~P. Kobayashi, Vertically self-organized inas
  quantum box islands on gaas(100), Phys. Rev. Lett. 75 (1995) 2542--2545.

\bibitem{Tarucha96}
S.~Tarucha, D.~G. Austing, T.~Honda, R.~J. van~der Hage, L.~P. Kouwenhoven,
  Shell filling and spin effects in a few electron quantum dot, Phys. Rev.
  Lett. 77 (1996) 3613.

\bibitem{Fujisawa98}
T.~Fujisawa, T.~H. Oosterkamp, W.~G. van~der Wiel, B.~W. Broer, R.~Aguado,
  S.~Tarucha, L.~P. Kouwenhoven, Spontaneous emission spectrum in double
  quantum dot devices, Science 282~(5390) (1998) 932--935.

\bibitem{Oosterkamp98}
T.~H. Oosterkamp, T.~Fujisawa, W.~G. van~der Wiel, K.~Ishibashi, R.~V. Hijman,
  S.~Tarucha, L.~P. Kouwenhoven, Microwave spectroscopy of a quantum-dot
  molecule, Science 395 (1998) 873--876.

\bibitem{Hayashi03}
T.~Hayashi, T.~Fujisawa, H.~D. Cheong, Y.~H. Jeong, Y.~Hirayama, Coherent
  manipulation of electronic states in a double quantum dot, Phys. Rev. Lett.
  91 (2003) 226804.

\bibitem{Shinkai09}
G.~Shinkai, T.~Hayashi, T.~Ota, T.~Fujisawa, Correlated coherent oscillations
  in coupled semiconductor charge qubits, Phys. Rev. Lett. 103~(5) (2009)
  056802.

\bibitem{Shinkai07}
G.~Shinkai, T.~Hayashi, Y.~Hirayama, T.~Fujisawa, Controlled resonant tunneling
  in a coupled double-quantum dot system, Appl. Phys. Lett. 90 (2007) 103116.

\bibitem{Frey12}
T.~Frey, P.~J. Leek, M.~Beck, A.~Blais, T.~Ihn, K.~Ensslin, A.~Wallraff, Dipole
  coupling of a double quantum dot to a microwave resonator, Phys. Rev. Lett.
  108 (2012) 046807.

\bibitem{Eriksson14}
Z.~Shi, C.~B. Simmons, D.~R. Ward, J.~R. Prance, X.~Wu, T.~S. Koh, J.~K.
  Gamble, D.~E. Savage, M.~G. Lagally, M.~Friesen, S.~N. Coppersmith, M.~A.
  Eriksson, Fast coherent manipulation of three-electron states in a double
  quantum dot, Nature Communications 5 (2014) 3020.

\bibitem{Fujisawa11}
T.~Fujisawa, G.~Shinkai, T.~Hayashi, T.~Ota, Multiple two-qubit operations for
  a coupled semiconductor charge qubit, Physica E 43 (2011) 730734.

\bibitem{Fanchini10}
F.~Fanchini, L.~K. Castelano, A.~O. Caldeira, Entanglement versus quantum
  discord in two coupled double quantum dots, New J. Phys. 12 (2010) 073009.

\bibitem{Emary09}
C.~Emary, Measuring the entanglement between double quantum dot charge qubits,
  Phys. Rev. B 80 (2009) 161309.

\bibitem{Hill97}
S.~Hill, W.~K. Wootters, Entanglement of a pair of quantum bits, Phys. Rev.
  Lett. 78 (1997) 5022.

\bibitem{Wootters98}
W.~K. Wootters, Entanglement of formation of an arbitrary state of two qubits,
  Phys. Rev. Lett. 80 (1998) 2245.

\bibitem{Nielsen}
M.~A. Nielsen, I.~L. Chuang, Quantum computation and Quantum information,
  Cambridge University Press, 2000.

\bibitem{CHSHpaper}
J.~F. Clauser, M.~A. Horne, A.~Shimony, R.~A. Holt, Proposed experiment to test
  local hidden-variable theories, Phys. Rev. Lett. 23 (1969) 880--884.

\end{thebibliography}

\end{document}